\documentclass[12pt,preprint]{aastex}

\usepackage{epsf}

\newcommand{\sect}[1]{\S\,\ref{#1}}

\newcommand{\be}{\begin{displaymath}}
\newcommand{\ee}{\end{displaymath}}
\newcommand{\bea}{\begin{eqnarray}}
\newcommand{\eea}{\end{eqnarray}}
\newcommand\msol{{M_{\odot}}}
\newcommand\mstar{{M}}
\newcommand\mr{{M}_r}

\newcommand\cs{\,cm$^2$\,s$^{-1}$}

\newcommand\dm{D_{\rm mix}}
\newcommand\dlt{\Delta\log\,T}

\shortauthors{Denissenkov et al.}
\shorttitle{Fluorine Variations as a Signature of Extra Mixing}

\begin{document}

\title{FLUORINE ABUNDANCE VARIATIONS AS A SIGNATURE OF ENHANCED EXTRA MIXING IN RED GIANTS OF THE GLOBULAR CLUSTER M4}

\author{Pavel A. Denissenkov\altaffilmark{1,2}, Marc Pinsonneault\altaffilmark{1}, \&
   Donald M. Terndrup\altaffilmark{1}}
\altaffiltext{1}{Department of Astronomy, The Ohio State University,
       140 West 18th Avenue, Columbus, OH 43210; dpa@astronomy.ohio-state.edu, pinsono@astronomy.ohio-state.edu,
   terndrup@astronomy.ohio-state.edu.}
\altaffiltext{2}{On leave from Sobolev Astronomical Institute of St. Petersburg State University,
   Universitetsky Pr. 28, Petrodvorets, 198504 St. Petersburg, Russia.}

\begin{abstract}
We show that enhanced extra mixing in low-mass red giants can result in
a fluorine abundance that is correlated with abundance variations of other
elements participating in H burning, such as C, N, O and Na. This finding is used to explain
the fluorine abundance variations recently found in bright red giants of the globular
cluster M4.
\end{abstract}


\section{Introduction}
\label{sec:intro}

In many galactic globular clusters, there are star-to-star abundance variations of
C, N, O, Na, Mg and Al, while the heavier elements show almost no dispersion
(see reviews by \citealt{d04b,gea04}, and references
therein). The signs of these variations,
their pairwise correlations, and the constancy of the sum of C+N+O in some clusters
(\citealt{p88,sea96,iea99}) indicate that we are most likely seeing the by-products of H burning in the CNO--,
NeNa--, and MgAl--cycles.
Regarding their production place and process, the most plausible alternatives proposed so far
include the so-called hot-bottom burning of hydrogen at the bottom of convective
envelope in intermediate-mass ($4\la\mstar/\msol\la 6$) asymptotic giant branch (IM-AGB) stars 
(\citealt{daea83,vea01}), and
the hydrogen shell burning in low-mass ($\mstar\la 2\,\msol$) red giant branch (RGB) stars
(\citealt{sm79,dd90,lh95}). The former case would imply primordial abundance variations
in the RGB stars where the abundance anomalies are observed, while the latter
explanation would require {\it in situ} mixing. The physical interpretation of these
results is therefore significant for understanding the origin of these anomalies.

Recently, \cite{sea05} have added $^{19}$F to the list of nuclides whose abundances
vary from star to star in globular clusters. They have found a deficit of the fluorine abundance
([F/Fe]\footnote{We use the standard spectroscopic notation: [A/B]\,=\,$\log\,(N({\rm
A})/N({\rm B}))-\log\,(N({\rm A})/N({\rm B}))_\odot$, where $N({\rm A})$ and 
$N({\rm B})$ are number densities of the nuclides A and B.}\,$<0$) 
anti-correlated with the abundance of Na in 7 bright red giants of the globular cluster M4.
The range of the fluorine abundance variation is $\sim$\,0.8\,dex.
Since $^{19}$F is effectively destroyed in the hot-bottom H burning
in the reaction $^{19}$F(p,$\alpha)^{16}$O, while low-mass
stars have so far been considered as its net producers (\citealt{fea92,jea92}),
\cite{sea05} have interpreted their measurements as an indication
that the star-to-star abundance variations in globular clusters originated in
the IM-AGB stars.

In this paper, we give an alternative interpretation of the new observational data
by demonstrating that low-mass RGB
stars could produce the F--Na anti-correlation in exactly the same way
as they contribute to the global O--Na anti-correlation
(\citealt{dv03,dea06}).

\section{IM-AGB and RGB Pollution Scenarios}
\label{sec:pollution}

In some globular clusters, the O--Na and Mg--Al anti-correlation and the N--Na
correlation have been found not only in evolved RGB stars but also in
main-sequence (MS) turn-off and early subgiant stars. Because these stars have interior temperatures
that are too low for the NeNa- and MgAl-cycles to work,
they must have abundance variations that predate the RGB phase; these are either present
{\it ab initio} or reflect accretion of material from IM-AGB stars
(\citealt{bea96,grea01,rc02,cea02,bea02,gea02,hea03}).

In IM-AGB stars, the envelope material processed in hot-bottom H burning is immediately
delivered to the interstellar medium via a strong stellar wind. On the other hand, in RGB stars
the H-burning shell is separated from the bottom of the convective envelope by a radiative zone.
However, extensive spectroscopic data on the surface chemical composition of low-mass RGB stars
as a function of luminosity, supported by appropriate stellar models, show that most of
these stars experience some extra mixing that connects the H burning shell
with the convective envelope (\citealt{sm79,ch94,ch95,grea00,dv03}). 
Near the RGB tip, before undergoing the core He-flash and
becoming a horizontal branch star,
a low-mass red giant loses a considerable amount of its envelope mass with a stellar wind
(\citealt{dcea96}), thus depositing products of
the H-shell burning to the interstellar medium. In globular clusters, both the IM-AGB stars
and the low-mass stars more massive than the present-day MS turn-off stars 
(i.e. $0.9\,\msol\la\mstar\la 2\,\msol$) have already
completed their lives. So, either of them could pollute the globular-cluster
interstellar medium with the ashes of H-burning.

Unfortunately, neither the AGB nor the RGB pollution scenario
can quantitatively reproduce all of the observed star-to-star abundance variations in globular clusters
simultaneously. For example, when destroying $^{16}$O in the IM-AGB stars
the hot-bottom burning depletes $^{24}$Mg even more (\citealt{dh03}).
It also keeps [C/Fe]
$\ga -0.5$ (\citealt{dw04}). 
Besides, the third dredge-up in AGB stars should increase the sum of C+N+O (\citealt{fea04}). 
None of these theoretical predictions is supported by observations. At present,
IM-AGB star models are being modified towards complying with constraints imposed
by the element abundance anomalies in globular clusters (e.g., \citealt{vda05}).

In low-mass RGB stars, extra mixing starts to manifest itself when the H-burning shell,
advancing in mass, erases the chemical composition discontinuity left behind by the bottom of
convective envelope at the end of the first dredge-up (\citealt{grea00,sh03}). At this moment, the evolution of 
red giants slows down for a while, which produces a prominent feature (the RGB bump) in the differential luminosity
functions of globular clusters (\citealt{zea99,rea03}). Therefore, extra mixing in low-mass stars is said to work
on the upper RGB, above this bump luminosity. Below the bump luminosity, on the lower RGB,
extra mixing is thought to be shielded from the H-burning shell by a gradient in the mean molecular weight
associated with the composition discontinuity 
(\citealt{sm79,chea98,dv03}) or to operate very slowly 
(\citealt{chea05,pea06}). 

The main problem of the RGB pollution
scenario is that for the majority of upper RGB stars, the observed pattern of surface
abundance anomalies can only be produced by {\it in situ} mixing that only penetrates
to the outer part of the H-shell, where the CN branch of the CNO--cycle is operating.
This is what \cite{dv03} have called ``canonical extra mixing''. They demonstrated
that its depth and rate (diffusion coefficient) can be
parameterized by any pair of correlated values within the close limits
specified by $\dlt\approx 0.19$ and $\dm\approx 4\times 10^8$\cs, to
$\dlt\approx 0.22$ and $\dm\approx 8\times 10^8$\cs.  Here, $\dlt$ is
the difference between the logarithms of temperature at the base of the
H-burning shell and at the maximum depth of extra mixing. Thus, canonical extra mixing
can be responsible for the evolutionary decline of [C/Fe] in upper RGB stars,
both in the field and in star clusters. It can also explain the decrease in surface
Li and $^3$He abundances with increased $L$, the strong reduction of the $^{12}$C/$^{13}$C isotopic ratio
and the slight increase in the N abundance, but it does not affect O, Na, Mg and Al.

However, if extra mixing in upper RGB stars penetrates the H-burning shell deeper
than in the canonical case, it could dredge up material deficient in O and enriched
in Na (\citealt{dd90}) and even in Al, under certain assumptions (\citealt{lh95,cea96,dt00,dw00}). 
\cite{dv03} have
proposed that in some upper RGB stars canonical extra mixing may be switched to
its enhanced mode with much faster and somewhat deeper mixing. If the extra mixing in
these stars is driven by differential rotation of their radiative zones then
such enhanced extra mixing could be caused by their spinning up as a result of
tidal synchronization in close binaries (\citealt{dea06}).

There is also an unexplored possibility that the depth and rate of 
canonical extra mixing do not remain
constant along the whole upper RGB but increase toward its tip.
This hypothesis is supported by the following arguments. Firstly, observations show that
in some globular clusters the anti-correlated abundance variations of C and N in red giants
become larger when stars approach the RGB tip. Moreover, the values of
[N/Fe]\,$\ga 1$ in some of these stars indicate the dredge-up of material in which not only C
but also a fraction of O has been converted into N (\citealt{sbh05}).
Secondly, at least in the globular cluster M13, the relative number of upper RGB stars with
the O--Na anti-correlation increases with luminosity (\citealt{jea05}). 
Note that all of the metal-poor field
stars used by \cite{dv03} to calibrate the parameters of canonical extra mixing have
$\log\,L/L_\odot\la 2.8$, i.e. they are located a magnitude below the RGB tip.
Hence, if canonical extra mixing got enhanced within the last magnitude of the upper RGB
then \cite{dv03} would not have seen it.

\section{$^{19}$F and Extra Mixing in Upper RGB Stars}

\subsection{Model Parameters for the M4 Red Giants}

The stellar evolution code, input physics and simple diffusion model of
extra mixing in upper RGB stars used here have been described by \cite{dea06}.
The code has since been developed to allow studies of the pre-MS and horizontal branch (HB) evolution.
As in many other similar codes, our zero-age HB models are constructed using information
about the internal structure of the RGB tip models
in which the core He-flash just sets in. 

\begin{figure}

\epsfxsize=12cm
\epsffile [20 160 480 680] {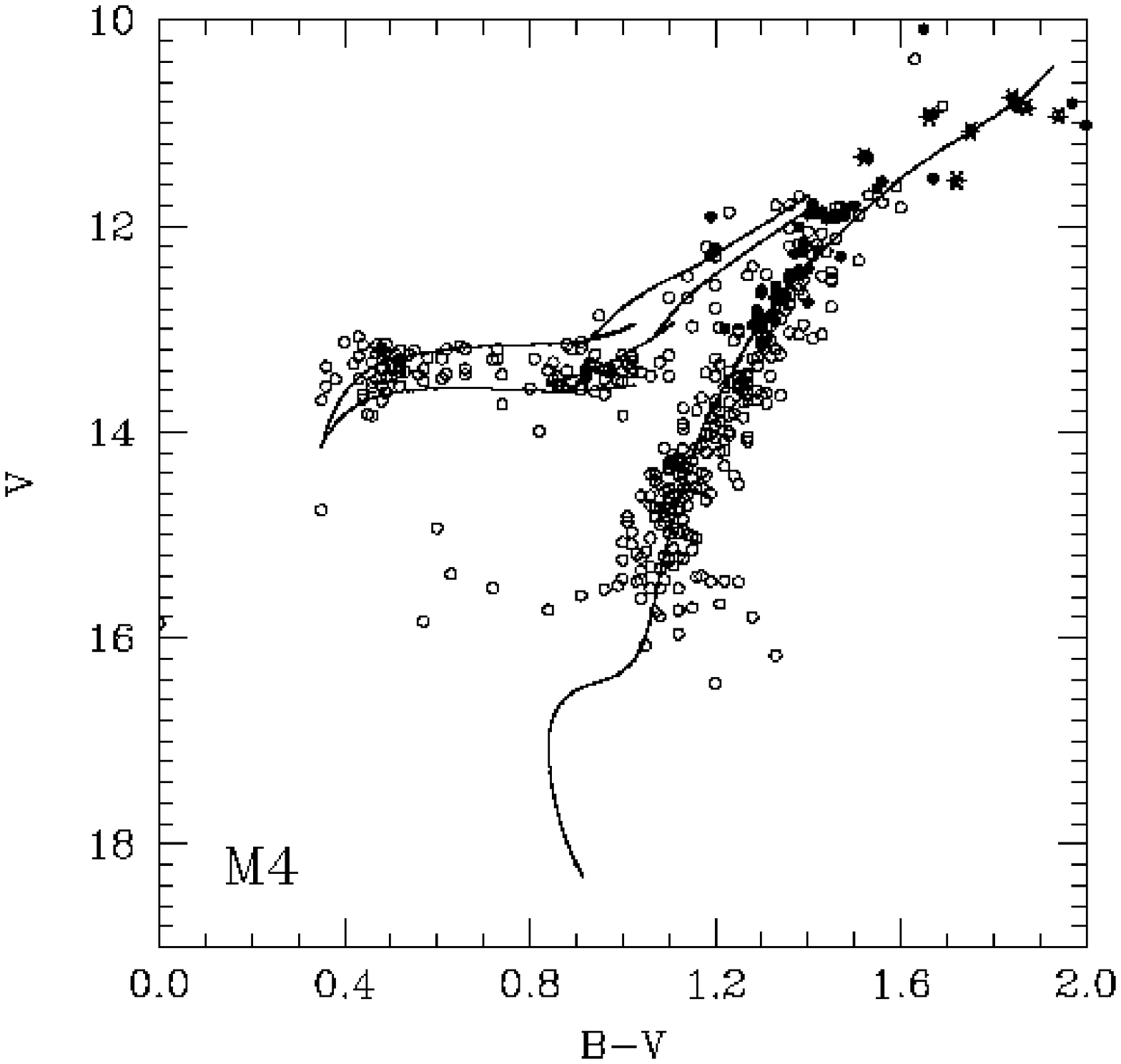}

\caption{CMD of evolved stars in M4 from the data of \cite{cr90}
         ({\it open circles}), \cite{sb05} ({\it filled circles})
         and \cite{sea05} ({\it asterisks}). {\it Curves} represent
         our evolutionary tracks for a model star with $\mstar = 0.83\,\msol$,
         $Z=0.002$ and $Y=0.24$ computed from the pre-MS to the RGB tip
         and, after that, from the zero-age HB (starting with $\mstar = 0.58\,\msol$ 
         and $\mstar = 0.62\,\msol$) to the beginning of the early AGB phase.
         The cluster reddening $E(B-V)=0.40$ and the distance modulus
         $(m-M)_{V,0}=12.83$ have been applied. These values are close to those
         used by \cite{aea97} and others.
        }
\label{fig:f1}
\end{figure}

We have found that for
a model star with the initial mass $\mstar = 0.83\,\msol$, helium and heavy-element
mass fractions $Y=0.24$ and $Z=0.002$ our RGB evolutionary track, zero-age HB and
HB evolutionary tracks fit the color-magnitude diagram (CMD) of evolved stars
in M4 reasonably well (Fig.~\ref{fig:f1}). Our adjusted stellar parameters give
a theoretical metallicity [Fe/H]\,$\approx -1.0$ and an age of evolved stars $\sim$\,14 Gyr.
In Fig.~\ref{fig:f1}, we have applied the cluster reddening $E(B-V)=0.40$ and the distance
modulus $(m-M)_{V,0}=12.83$. Our adopted parameters are in good agreement with those
used by \cite{aea97} who fitted the M4 main-sequence CMD with
a 14 Gyr isochrone of \cite{bv92} for $Y=0.2388$, [Fe/H]\,$=-1.03$, $E(B-V)=0.42$
and $(m-M)_{V,0}=12.80$. The spectroscopic metallicity of M4 is [Fe/H]\,$= -1.18$
(\citealt{iea99}).

\subsection{$^{19}$F and Canonical Extra Mixing}

\begin{figure}

\epsfxsize=12cm
\epsffile [20 160 480 680] {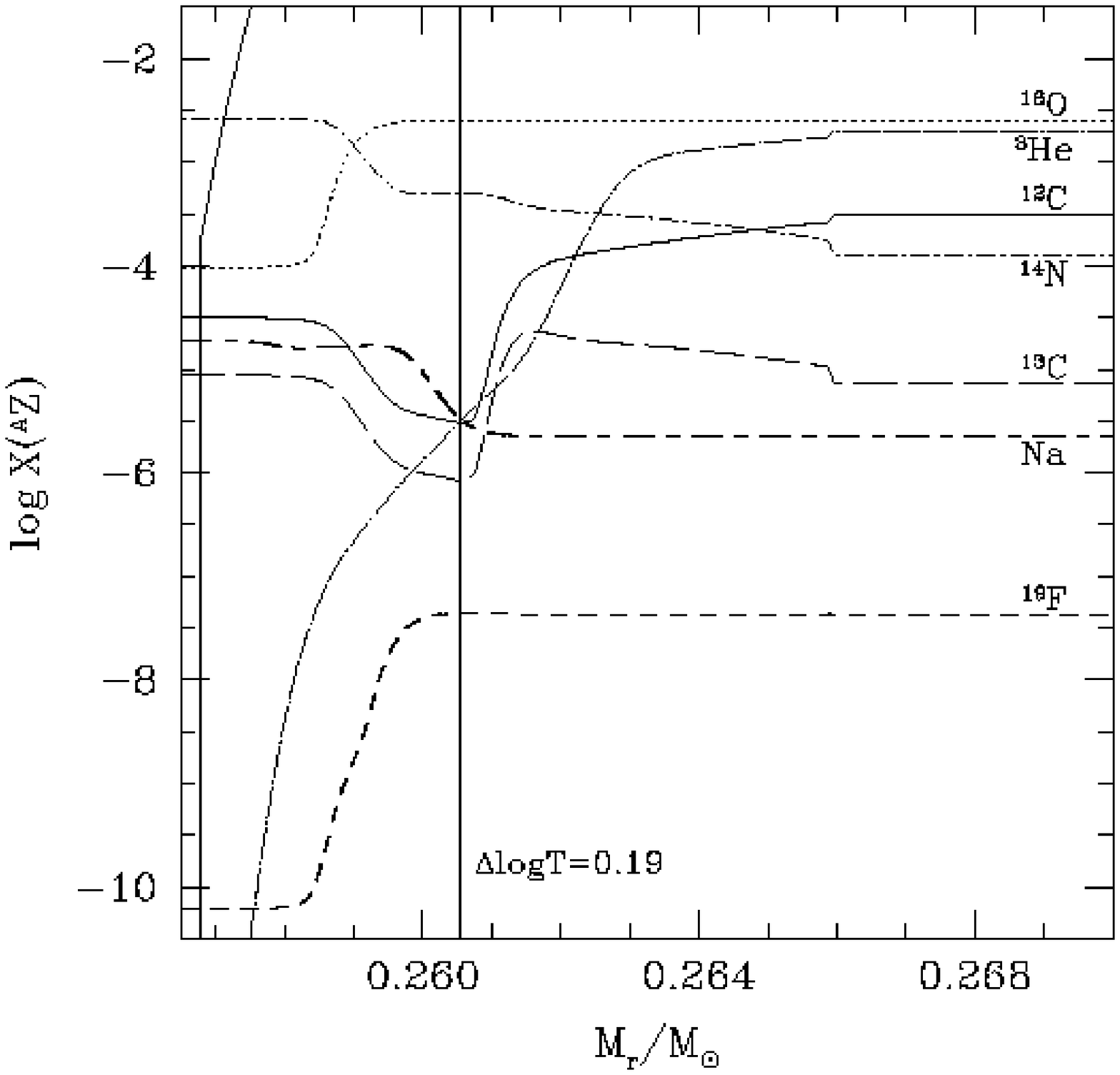}
\caption{Abundance profiles of some nuclides near the H burning shell
         (its bottom coincides with the vertical solid line on the left
         showing the H abundance profile) in a bump luminosity $0.83\,\msol$ model.
         The vertical line marks the depth characteristic of
         canonical extra mixing (\citealt{dv03}). It is seen that
         such shallow mixing will not change the surface fluorine
         abundance.
        }
\label{fig:f2}
\end{figure}

Abundance profiles of some nuclides participating in the H-shell burning, including $^{19}$F,
in a bump luminosity model star are plotted in Fig.~\ref{fig:f2}.
The vertical line marked $\dlt = 0.19$ is shown at a depth characteristic of canonical extra mixing.
The nearly vertical line at the left of the plot shows the profile for H;
this coincides with the lower edge of the H-burning shell. Canonical extra mixing
occurs when the H-burning shell erases the composition discontinuity at
the mass coordinate $\mr\approx 0.266\,\msol$. Most or all stars more luminous
than this would show a reduced abundance of $^{12}$C (\citealt{dv03}). 
However, we see that canonical extra mixing
will not change the surface abundance of $^{19}$F.

\subsection{$^{19}$F and Enhanced Extra Mixing}

Note that all of the 7 red giants studied by \cite{sea05} are located within
one magnitude of the RGB tip ({\it asterisks} in Fig.~\ref{fig:f1}). 
As we have already mentioned in \sect{sec:pollution}, there are some
arguments supporting the idea that at least the depth of canonical extra mixing increases
in the vicinity of the RGB tip. In particular, the decline of [C/Fe] becomes stronger
within the last magnitude of the RGB, which is accompanied by extremely high N abundances
([N/Fe]\,$\ga 1$) supposedly signifying the dredge-up of material with O partially processed into N
(\citealt{sbh05}). \cite{sea05} claim that they have found a similar decline of the C abundance
$A(^{12}\mbox{\rm C})\equiv\log\,N(\mbox{\rm C})/N(\mbox{H}) + 12$ with the absolute bolometric
magnitude in their M4 stars ({\it circles with errorbars} in the left panel in Fig.~\ref{fig:f3}).

\begin{figure}

\epsfxsize=12cm
\epsffile [20 160 480 680] {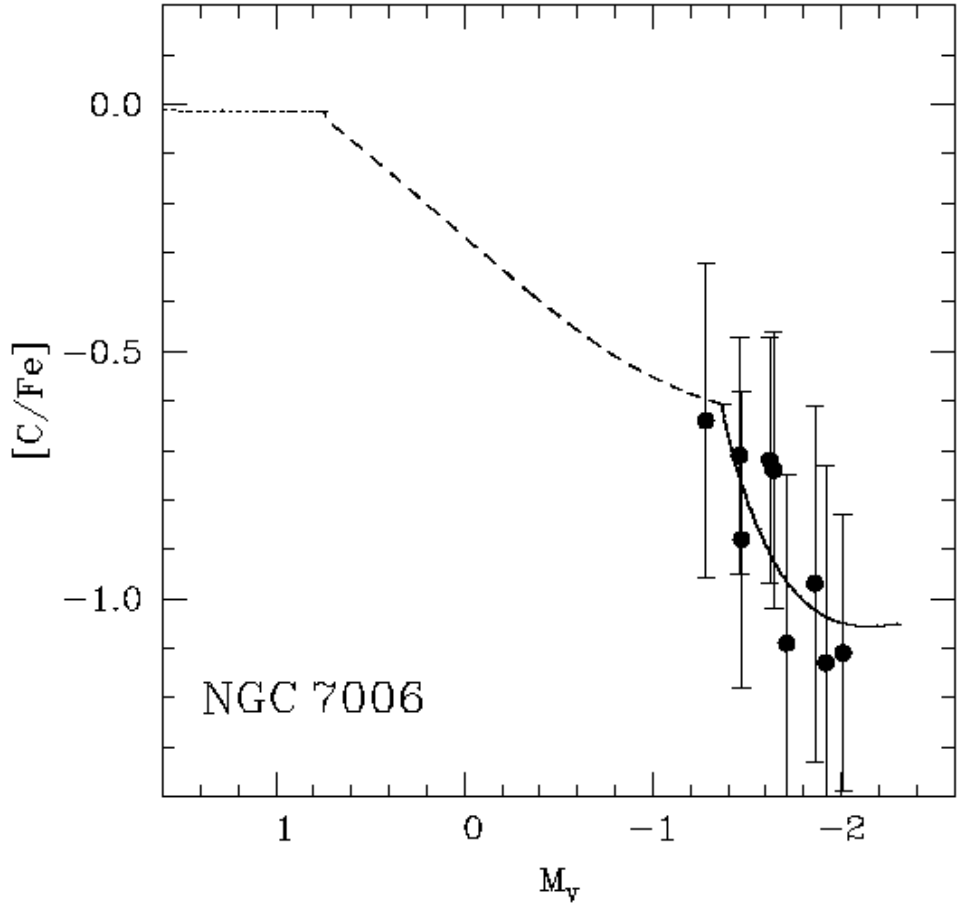}
\caption{Evolutionary decline of [C/Fe] with $M_V$ caused by   
         canonical extra mixing with $\dlt = 0.19$ and $\log\,\dm =8.6$
         ({\it dashed curve}) followed by enhanced extra mixing with $\dlt = 0.05$
         and $\log\,\dm = 9.6$ ({\it solid curve}). {\it Dotted line} gives
         the post-first-dredge-up value of [C/Fe]. {\it Circles with errorbars} are 
         observational data for bright red giants of the globular
         cluster NGC\,7006 (\citealt{sea05}).
        }
\label{fig:f5}
\end{figure}

In order to test the hypothesis that canonical extra mixing gets enhanced near the RGB tip,
we have proceeded as follows. First, starting with a model at
$\log L/L_\odot = 1.75$ (this value is slightly less than the bump luminosity;
it corresponds to $M_{\rm bol}=0.37$, $M_V=0.72$, and $V=13.55$),
we have computed its evolution up to $\log L/L_\odot = 2.74$ ($M_{\rm bol}=-2.10$, 
$M_V=-1.37$, and $V=11.46$). In this computation, the depth and rate of extra mixing
had their canonical values $\dlt = 0.19$ (Fig.~\ref{fig:f2})
and $\dm = 4\times 10^8$\,\cs, or $\log\dm = 8.6$. Abundance profiles in the radiative zone
in the final model are shown in Fig.~\ref{fig:f4}. After that, we have continued
our stellar evolution computations toward the RGB tip with increased
depth and rate of extra mixing. The parameter $\dlt$ has been reduced to 0.05, while
for $\log\dm$ we have considered the following higher values: 9.2, 9.3, 9.4, 9.5, and 9.6.
Already looking at Fig.~\ref{fig:f4}, one can conclude
that such enhanced extra mixing will dredge up
material in which deficits of $^{16}$O and $^{19}$F should correlate with overabundances of
N and Na. Curves in left panel in Fig.~\ref{fig:f3} demonstrate how
the surface C abundance declines with $M_{\rm bol}$ in our models
with enhanced extra mixing. In Fig.~\ref{fig:f5}, a combined evolution of
[C/Fe] is shown as a function of $M_V$ for the post-first-dredge-up ({\it dotted line}),
canonical extra mixing ({\it dashed curve}), and enhanced extra mixing ({\it solid curve})
phases. Here, points with errorbars are data for bright red giants of the globular
cluster NGC\,7006 from the work of \cite{sea05}, the 4 brightest of them having
[N/Fe]\,$> 1.3$ (given the large uncertainties of the [C/Fe] data in NGC\,7006,
the difference in metallicities [Fe/H]$_{\mbox{\small NGC\,7006}}-$\,[Fe/H]$_{\mbox{\small M4}}\approx -0.5$
is unimportant for the comparison).

\begin{figure}

\epsfxsize=12cm
\epsffile [20 160 480 680] {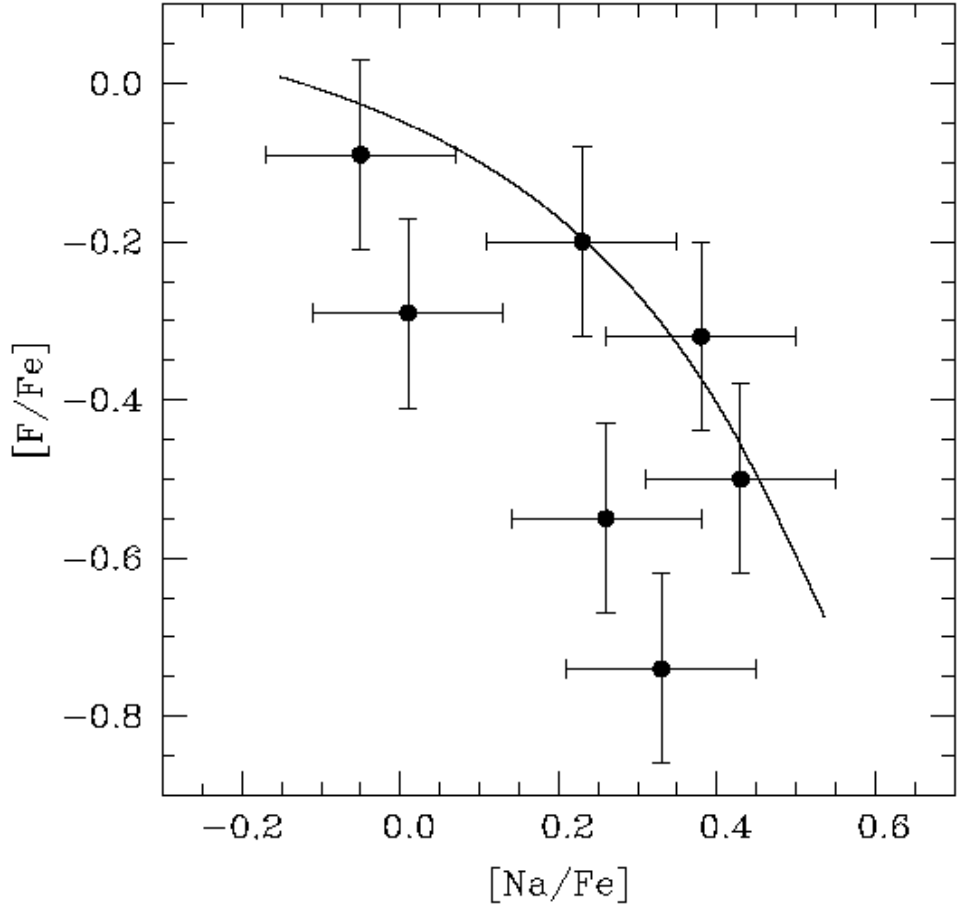}
\caption{The Na--F anti-correlation revealed by \cite{sea05} in 7 bright 
         red giants of the globular cluster M4 ({\it circles} with errorbars).
         {\it Curve} shows our theoretical dependence of [F/Fe] on [Na/Fe]
         obtained as a result of enhanced extra mixing with $\dlt = 0.05$ and
         $\log\,\dm = 9.6$ that is assumed to replace canonical exra mixing
         when stars approach the RGB tip (for details, see text).
        }
\label{fig:f6}
\end{figure}

Interestingly, if we take the fluorine abundances measured by \cite{sea05}
in the M4 red giants at their face values and plot them as a function of $M_{\rm bol}$,
we find a correlation of $A(^{19}\mbox{\rm F})$ 
with $M_{\rm bol}$ ({\it circles with errorbars} in right panel
in Fig.~\ref{fig:f3}) that looks none the worse than the correlation between
$A(^{12}\mbox{\rm C})$ and $M_{\rm bol}$ (left panel). Even more interestingly is that
the curves in the right panel representing results of our evolutionary computations
with enhanced extra mixing could nicely reproduce that correlation if it were real.
We also find a theoretical anti-correlation between [Na/Fe] and [F/Fe] that
is qualitatively similar to the observational one revealed by \cite{sea05} 
(Fig.~\ref{fig:f6}).

Therefore, alternatively to the IM-AGB pollution scenario, we propose that
the fluorine abundance variations in the red giants of the globular cluster M4
may actually be a signature of enhanced extra mixing in low-mass RGB stars.
Moreover, we strongly encourage spectroscopists to increase the size of the sample of
bright red giants with known fluorine abundances in globular clusters because
if $A(^{19}\mbox{\rm F})$ indeed correlates with $M_{\rm bol}$ this will be
a direct evidence of enhanced extra mixing in these stars.
On the other hand, if the fluorine abundance is not found to decline with
increasing luminosity near the RGB tip this will not necessarily reject our hypothesis that its deficit
is due to enhanced extra mixing in upper RGB stars.
There will still remain a possibility that the star-to-star abundance
variations of $^{19}$F, like those making up the O--Na anti-correlation,
were produced by enhanced extra mixing in the low-mass red giants
that had been slightly more massive than the present-day MS turn-off stars in
globular clusters and that had completed their lives in the past (the RGB pollution
scenario). 

We have also examined the possibility that the F/Na anticorrelation
could have been produced by larger-than-expected temperature errors.
The F abundances were computed using the temperatures determined by
\cite{iea99}, who obtained them by measuring the line depth
ratios of temperature-sensitive species following \cite{gj91}.
The stated random errors in the temperature were about 50 K.
Using the sensitivity of the abundances to temperature, we find that the
dispersion in the F or Na abundances could have been produced if the
random errors in temperature were as large as 125 K.  Since both lines
have the same sense of temperature dependence, however, such large
random errors would produce an erroneous correlation between the F and
Na abundances, instead of the observed anticorrelation.

Note that for a Salpeter initial mass function the total mass lost by
those red giants before they arrived at the zero-age HB is comparable to the mass
delivered to the interstellar medium by the IM-AGB stars (\citealt{d04a}). This estimate
assumes that every upper RGB star loses $0.2\,\msol$ before it undergoes the core He-flash.
For comparison, our HB model stars with $(B-V)<0.7$ on the CMD of the globular cluster M4 in Fig.~\ref{fig:f1} all 
have $\mstar < 0.61\msol$, i.e. each of them has lost more than $0.22\,\msol$.
It is far from clear, however, that the problem of depositing the lost mass from
either AGB or RGB stars onto the other stars in the system after they reached the MS
has been satisfactorily addressed. For this reason, determining the contribution of
direct mixing on the observed abundances is important.

\section{Conclusion}

\begin{figure}

\epsfxsize=12cm
\epsffile [20 160 480 680] {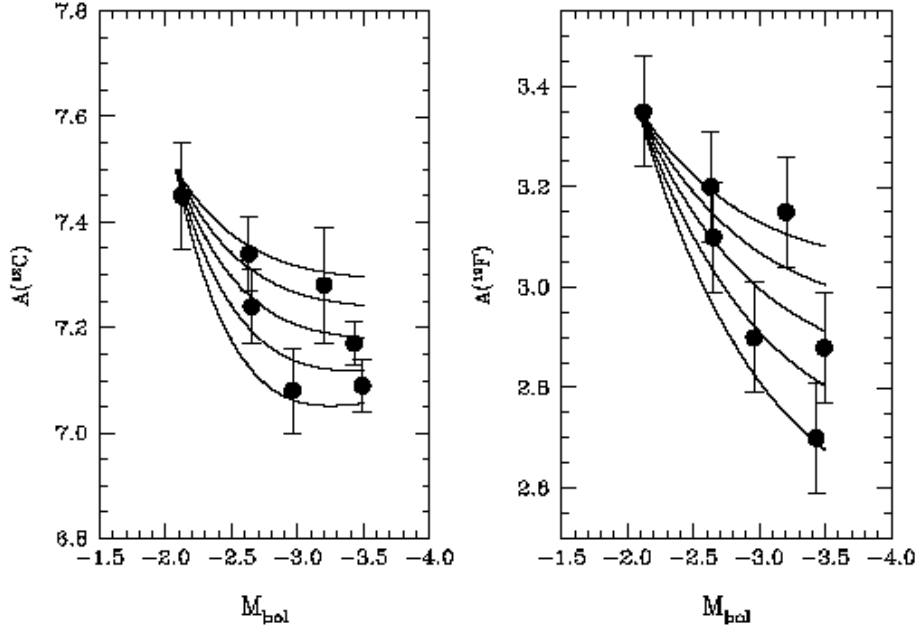}
\caption{The correlation between the surface C abundance and $M_{\rm bol}$
         claimed by \cite{sea05} to exist in their observed bright
         red giants of the globular cluster M4 ({\it circles} with errorbars in
         the left panel) and a possible correlation between the fluorine abundance and
         $M_{\rm bol}$ revealed by us in their published data (the right panel).
         {\it Curves} represent results of our evolutionary computations
         with enhanced extra mixing near the RGB tip for the mixing depth $\dlt = 0.05$
         (Fig.~\ref{fig:f4}) and rates $\log\,\dm = 9.6$, 9.5, 9.4, 9.3,
         and 9.2 (from the bottom to the top curve).
        }
\label{fig:f3}
\end{figure}

\begin{figure}

\epsfxsize=12cm
\epsffile [20 160 480 680] {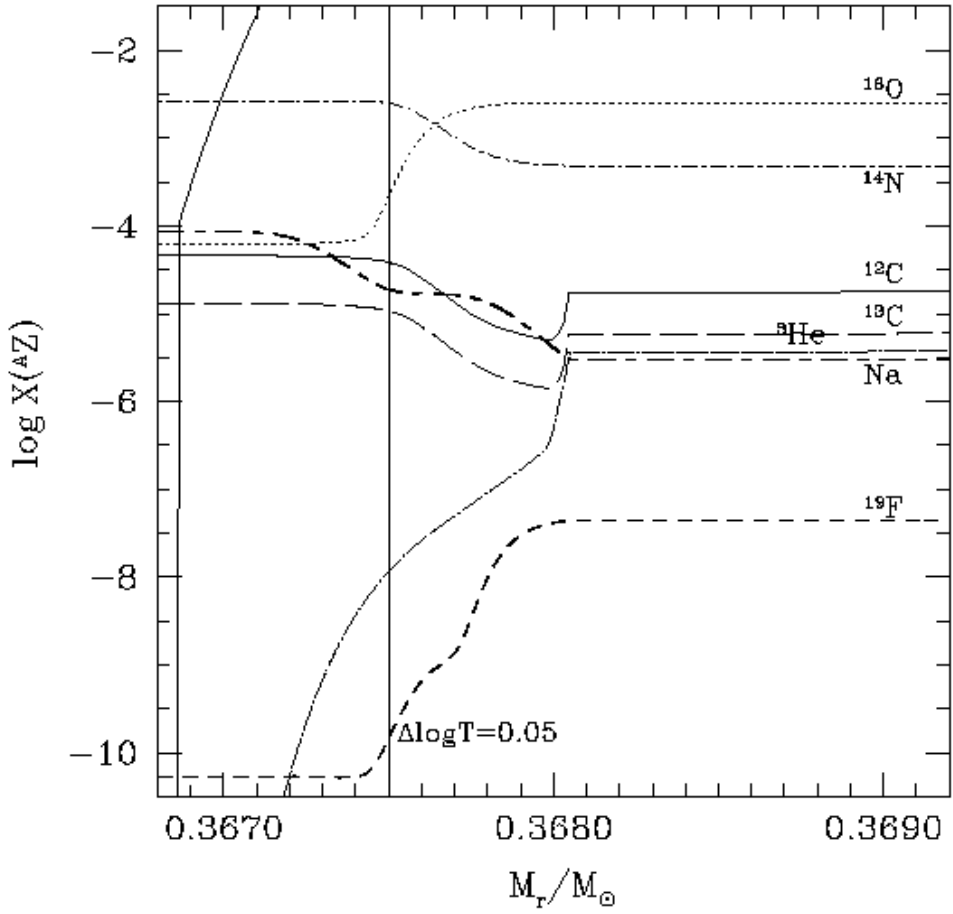}
\caption{The same as in Fig.~\ref{fig:f2} but for a model near
         the RGB tip. For enhanced extra mixing with the depth shown
         with the solid line, the surface fluorine abundance
         undergoes considerable changes (see the right panel in
         Fig.~\ref{fig:f3}).
        }
\label{fig:f4}
\end{figure}

In this work, we have shown that the reduced abundances of $^{19}$F (i.e. [F/Fe]\,$<0$)
anti-correlated with [Na/Fe] found by \cite{sea05} in the red giants of M4 do not rule out the RGB
pollution scenario in which canonical extra mixing switches to its enhanced mode.
The enhanced extra mixing could result from tidal spin-up of upper RGB stars in close binaries (\citealt{dea06}).
Another possibility is that canonical extra mixing
gets enhanced toward the RGB tip due to some internal physical processes in single stars.
We emphasize that there is observational evidence of this, one of which could be
a correlation of $A(^{19}\mbox{\rm F})$ with $M_{\rm bol}$ like that plotted in
the right panel in Fig.~\ref{fig:f3}, provided it is confirmed in
future spectroscopic observations.
At present the reduced fluorine abundance cannot be considered as a strong argument in favor of
the IM-AGB pollution scenario as the only plausible interpretation of
star-to-star abundance variations in globular clusters. 
On the contrary, if $A(^{19}\mbox{\rm F})$ is correlated with $M_{\rm bol}$ in low-mass stars
near the RGB tip, this will
support the original hypothesis of \cite{dd90} that
extra mixing in upper RGB stars can penetrate the H burning shell deep enough
to dredge-up material enriched not only in N but also in Na, and deficient
not only in C but also in O and F.
A small number of stars in the fast evolutionary phase, at the RGB tip, could explain 
the absence of field upper RGB stars in which the anti-correlated O and Na 
surface abundances are produced and dredged up {\it in situ} (\citealt{grea00}).

\acknowledgements
We acknowledge support from NASA grant \#\,NNG05 GG20G.


\clearpage



\clearpage



\clearpage



\clearpage



\clearpage



\clearpage


\end{document}